\documentclass[pof,aps,twocolumn]{revtex4-1}
\usepackage{graphicx} 
\usepackage[usenames]{color}
\usepackage{amsmath,amssymb}
\usepackage{gensymb}
\usepackage{natbib}
\usepackage{xcolor}
\def\strutdepth{\dp\strutbox}
\def\nw#1{\strut\vadjust{\kern-\strutdepth\vtop to0pt{\vss\hbox to\hsize
{\hskip\hsize\hskip5pt$\leftarrow$\hss\strut}}}{\em #1}}

\begin{document}

\title{Elastocapillary instability under partial wetting conditions: bending versus buckling}

\author{Bruno Andreotti$^1$, Antonin Marchand$^1$,  Siddhartha Das$^2$ and Jacco H. Snoeijer$^2$}
\affiliation{
$^{1}$Physique et M\'ecanique des Milieux H\'et\'erog\`enes, UMR
7636 ESPCI -CNRS, Univ. Paris-Diderot, 10 rue Vauquelin, 75005, Paris\\
$^{2}$Physics of Fluids Group and J. M. Burgers Centre for Fluid Dynamics,
University of Twente, P.O. Box 217, 7500 AE Enschede, The Netherlands.
}

\date{\today}%

\begin{abstract}
The elastocapillary instability of a flexible plate plunged in a liquid bath is analysed theoretically. We show that the plate can bend due to two separate destabilizing mechanisms, when the liquid is partially wetting the solid. For contact angles $\theta_e > \pi/2$, the capillary forces acting tangential to the surface are compressing the plate and can induce a classical buckling instability. However, a second mechanism appears due to capillary forces normal to surface. These induce a destabilizing torque that tends to bend the plate for any value of the contact angle $\theta_e > 0$. We denote these mechanisms as ``buckling'' and ``bending'' respectively and identify the two corresponding dimensionless parameters that govern the elastocapillary stability. The onset of instability is determined analytically and the different bifurcation scenarios are worked out for experimentally relevant conditions. 
\end{abstract}

\maketitle

\section{Introduction}
\begin{figure}[t!]
\includegraphics{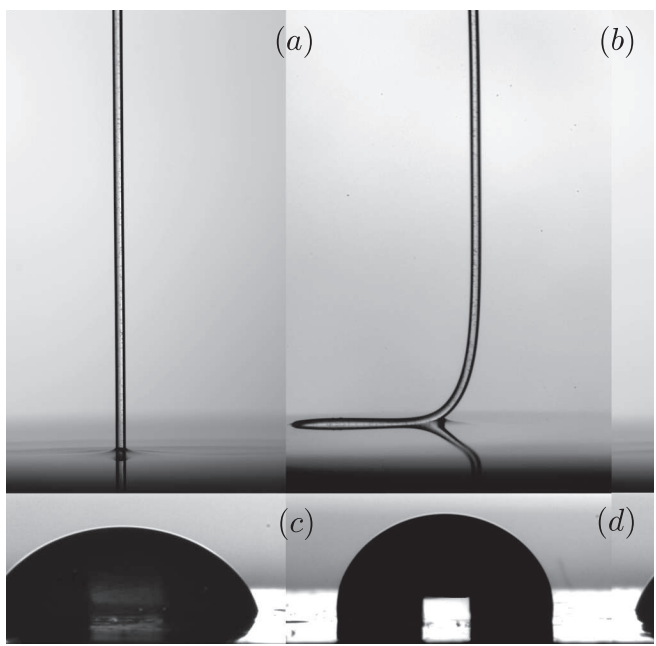}
\caption{(a,b) Photographs of an elastomeric wire of radius $R=300\,\mu $m and density $\rho=1.02 \cdot 10^3 \, {\rm kg/m^3}$ brought into contact with a mixture of ethanol and water. From (a) to (b), the contact angle is increased continuously by decreasing the concentration of ethyl alcohol. The wire exhibits a buckling instability above a critical contact angle. (c,d) Independent determination of the contact angle, using drops of the same mixture on a plane substrate made of the same elastomer as in (a,b). On the left, the advancing contact angle is 60$^{\circ}$ with a 50\% volumic solution of ethyl alcohol in water. On the right, the advancing contact angle is 95$^{\circ}$ with a 5\% volumic solution of ethyl alcohol in water.}
\label{fig:bending}
\end{figure}

Water-walking arthropods like water striders are able to float, despite their density, thanks to surface tension forces \cite{BushARFM}. Their superhydrophobic legs are submitted to a repulsive force along the contact line where the three phases (liquid, vapour and solid) meet. As the legs are long and flexible, they deform under these capillary forces~\cite{ParkJFM}. Figure~\ref{fig:bending} shows an experiment performed with extremely long artificial legs made of a soft solid, which is plunged into a liquid bath. One observes an elastocapillary instability that is triggered by increasing the contact angle $\theta_e$: the initially immersed solid is pushed out of the liquid to the free surface, whenever $\theta_e$ is sufficiently large.

In the limiting case of complete wetting, i.e. $\theta_e=0$, a compressive force is exerted on an elastic rod initially immersed in a liquid when its end pierces the liquid interface \cite{ChakrapaniPNAS,NeukJMPS,ChiodiEPL}. Such a rod buckles when the compressive force exceeds EulerÕs critical load. Consider the case of a plate of thickness $H$ much smaller than its length $L$ and its width $W$. It is submitted to a capillary force equal to the  water surface tension $\gamma_{LV}$ times the perimeter $\simeq 2W$. The critical force is equal to  $(\pi/2)^2BW/L^2$, where $B$ is the bending stiffness, which can be expressed as $B=EI/W$, where $E$ it the Young's modulus, $I$ the moment of inertia and $W$ the width of the plate. 
%For the plate one finds $B=EH^3/12$. 
Therefore, buckling occurs if the plate is longer than a critical length $L_{\rm cr}=(\pi/2)\;\sqrt{B/(2\gamma_{LV})}$. It is proportional to the elasto-capillary length
\begin{equation}\label{eq:bico}
\ell_{EC} = \left( \frac{B}{\gamma_{LV}} \right)^{1/2},
\end{equation}
which is the length scale controlling a large class of elastocapillary problems \cite{BicoNATURE,NeukJMPS,PyPRL,BoudPRE,PyEPJST,RomanJPCM,ChiodiEPL,HurePRL,HonsAPL}.

\begin{figure}[t]
\begin{center}
\includegraphics{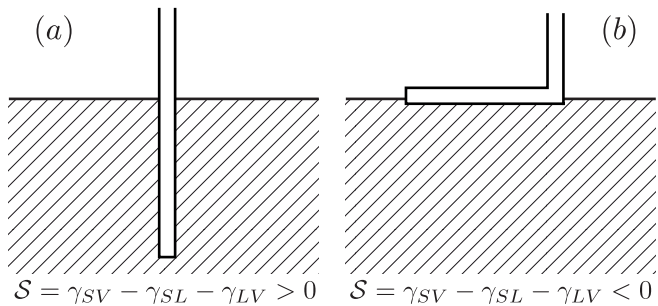}
\caption{\label{fig:energy} Capillary energy of a partially immersed plate, in the absence of gravity and elasticity. (a) Two sides of the plate are immersed, representing an energy of $2\gamma_{SL}$ per immersed length. (b) When bending towards the surface, one of the sides is no longer wetted and part of the liquid-vapor interface is covered. The associated energy per length is $\gamma_{SV}+\gamma_{SL}-\gamma_{LV}$. The energy difference between (b) and (a) equals the spreading parameter $\mathcal{S}=\gamma_{SV}-\gamma_{SL}-\gamma_{LV}$. For the partially wetting case $\mathcal{S}<0$, so that state (b) is energetically more favorable than state (a).}
\end{center}
\end{figure}

One may wonder if the instability observed in Fig.~\ref{fig:bending} is of the same physical nature. Indeed, one can expect a buckling instability if the contact angle $\theta_e$ is larger than $\pi/2$. Namely, the total downward force that the reservoir exerts on the solid is proportional to $\gamma_{LV}\cos \theta_e$, and hence changes from ``stretching'' to ``compressing'' when the contact angle exceeds $\pi/2$. However, there is a second mechanism that can lead to elastic deformations. Figure~\ref{fig:energy} compares the capillary energy of an extremely flexible object that either remains vertical or floats on the free surface of the liquid. The free energy difference is proportional to the spreading parameter $\mathcal{S}=\gamma_{SV}-\gamma_{SL}-\gamma_{LV}$, where $\gamma_{SV}$, $\gamma_{SL}$ and $\gamma_{LV}$ are respectively the surface tensions of the solid/vapour, solid/liquid and liquid/vapour interfaces. As a consequence, bending is favourable under partial wetting conditions, ${\cal S}<0$, whatever the value of the contact angle $\theta_e$. This is manifestly different from the buckling instability, which can only occur for $\theta_e > \pi/2$. Indeed, the mechanism for instability is not the vertically compressing force, but is a bending induced by the capillary torque exerted near the contact line \cite{ParkJFM}.

In this paper, we investigate theoretically the instability of an elastic plate plunged in a liquid of the same density --~to avoid buoyancy. In the light of former studies, we wish to address different issues. What are the mechanisms for instability: bending, buckling or a combination of the two? What is the relevant length $L$ for the instability in the situation of Fig.~\ref{fig:bending}, where the plate can be supposed to be infinite? What are the parameters controlling the instability? 

To illustrate the two mechanisms of elastocapillary instability, bending and buckling, we consider a long elastic plate that is hanging freely under the influence of gravity. We assume the thickness to be sufficiently small to allow for a thin plate elastic description. The bottom of the plate is brought into contact with a liquid reservoir that partially wets the liquid, with an equilibrium contact angle $\theta_e$. To reveal the effect of surface wettability, we focus on the case where both sides of the plate are wetted by the same angle. This is fundamentally different from the situation prior to piercing of a rod through a meniscus~\cite{ParkJFM,ChakrapaniPNAS,NeukJMPS,ChiodiEPL}, for which one of the contact lines is pinned to the edge of the solid -- in that case the contact angle can attain any value. Our goal is to compute the shape of the plate and to analytically determine the threshold of instability for different $\theta_e$.  

Our main finding is that the elastocapillary instability can occur even when $\theta_e < \pi/2$, which is the regime where the capillary forces are pulling on the plate and no ``buckling'' is to be expected. Indeed, this is due to the bending mechanism described in Fig.~\ref{fig:energy}, due to the partial wettability of the substrate. In general, the threshold of instability depends on two dimensionless parameters that can be associated to bending and buckling respectively. Interestingly, the bending parameter is not only determined by the elastocapillary length $\ell_{EC}$, but also involves the characteristic size of the meniscus. 

The paper is organized as follows. We first formulate the elastocapillary problem and identify the relevant dimensionless quantities in Sec.~\ref{sec:formulation}. In Sec.~\ref{sec:instability} we analytically determine the threshold of instability by linear analysis and numerically compute the nonlinear bifurcation diagrams. The results are interpreted in experimental context in Sec.~\ref{sec:exp}, where we also discuss the influence of contact angle hysteresis. The paper concludes with a discussion on the distribution of capillary forces in Sec.~\ref{sec:discussion}.

\section{Elastocapillary formulation}\label{sec:formulation}

\begin{figure}[t]
\begin{center}
\includegraphics{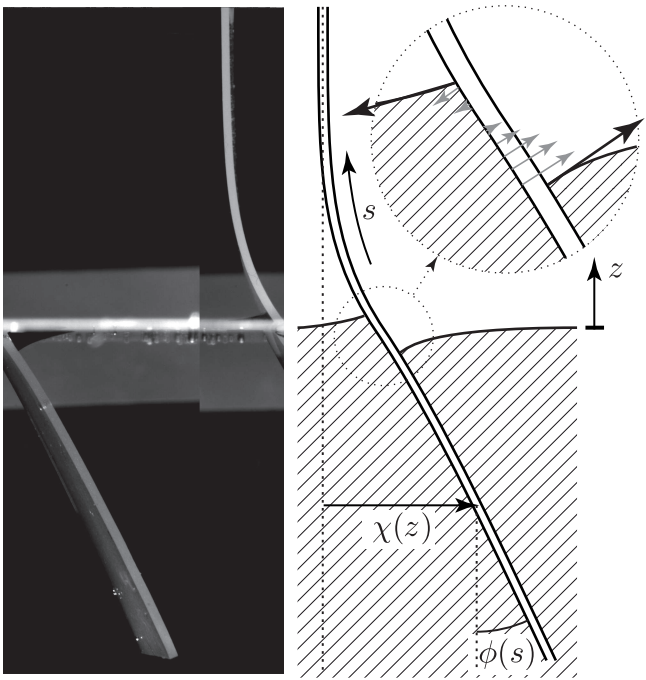}
\caption{\label{fig:analysis} Left: Photograph of an elastomeric plate partially bent by capillary forces. We show that such bending to a stable, finite angle is only possible due to contact angle hysteresis. The plate thickness is $H=0.8\,$mm, its Young modulus is $E=66\,$ kPa and its density is $1.02 \cdot 10^3 \, {\rm kg/m^3}$. As not particularly clean tap water is used,  the surface tension is around $\gamma=0.05~{\rm N/m}$. Right: Definitions of the vertical coordinate $z$ and curvilinear coordinate $s$. The deformation of the plate is characterized by the deflection $\chi(z)$ or local angle $\phi(s)$. The characteristic curvature of the plate is $\ell_\chi^{-1}$. Inset: sketch of the resultant capillary forces in the meniscus region. At the contact lines there are pulling forces along the liquid-vapor interface of magnitude $\gamma$. In addition, the hydrostatic pressure pulls or pushes on the plate depending on the level with respect to bath. The scale of the meniscus is $\ell_\gamma$.}
\end{center}
\end{figure}

The strategy of the calculation is to separately treat the portion of the plate that is outside the bath and the meniscus region near the contact line -- see Fig.~\ref{fig:analysis}. For simplicity we assume that the plate and fluid are density matched, or equivalently, that the bottom of the plate reaches only just below the surface. In practice, we find that the characteristic length of the plate outside the reservoir is significantly larger than the capillary length that sets the size of the meniscus. This means that we can consider the forces and torques induced by the meniscus (Sec.~\ref{sec:meniscus}) as a boundary condition for the dry part of the plate (Sec.~\ref{sec:outside}). The dimensionless equations and boundary condition are then summarized in Sec.~\ref{sec:dimensionless}.

\subsection{The plate outside the reservoir}\label{sec:outside}

Given that the plate is very thin we can describe the shape by a line that we parametrize by its angle $\phi(s)$ with respect to the vertical direction (Fig.~\ref{fig:analysis}). We use a curvilinear coordinate $s$ that has $s=0$ at the level of the bath, and the curvature is $\kappa=d\phi/ds$. In cartesian coordinates, we use the parametrisation $x= \chi(z)$, where $z$ is orientated upward and $z=0$ corresponds to the level of the liquid reservoir. The relation between the two representations is: $d\chi/ds=\sin \phi$ and $dz/ds=\cos \phi$. We consider a very long plate that, due to gravity, follows the boundary condition $\phi(\infty)=0$, cf. Fig.~\ref{fig:analysis}. 

Since away from the meniscus there are no forces applied to the plate, the shape can be computed from the elastica equations~\cite{Timoshenko,Landau}:
\begin{equation}\label{eq:el1}
B \phi'' = F_z \sin \phi.
\end{equation}
Here, $B$ is the bending stiffness, which reflects the internal elastic torque (per unit width) due to a curvature $\phi'$. For a thin plate of thickness $H$ and elastic modulus $E$ one finds $B=EH^3/12$. $F_z$ is the vertical component of the force (per unit plate width) on a cross-section of the plate and we will find below that $F_x=0$. The torque balance then reads:
\begin{equation}\label{eq:Mi}
{\cal T}_i = - B \phi' .
\end{equation}
This internal force moment is exerted by the upper portion of the plate on the lower portion of the plate. Note that for the situation in Fig.~\ref{fig:analysis} the curvature is negative, $\phi' < 0$. 

The force $F_z$ on a cross-section consists of two contributions, due to gravity along the plate and due to surface tension at the meniscus boundary. At a location $s$ along the plate, the vertical force due to gravity is simply the weight below $s$ , i.e. $\rho g Hs$ per unit width of the plate. As we assume the plate to have the same density as the liquid, or equivalently that the bottom dips just below the surface, we only take into account the portion of the plate that is outside the reservoir.

\begin{figure}[t]
\begin{center}
\includegraphics{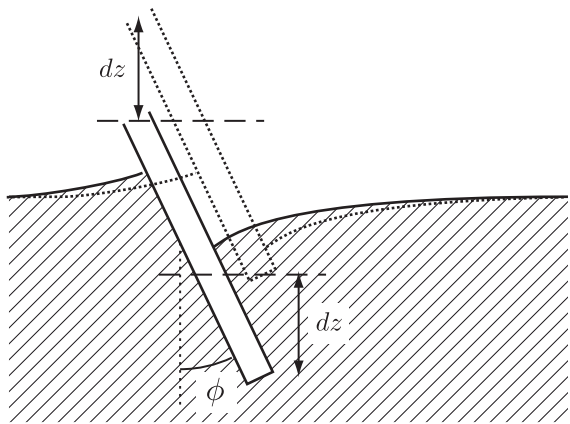}
\caption{\label{fig:virtual} The resultant vertical force on the plate can be obtained from the change in surface free energies due to a virtual displacement $dz$. See text for details.}
\end{center}
\end{figure}

The capillary forces exerted in the meniscus region can be obtained by the virtual work principle, as sketched in Fig.~\ref{fig:virtual}. Let us consider the left side of the plate. Moving the plate vertically by $dz$, one changes the horizontal position of the contact line by $dz \tan \phi(0)$, leading to an increase of the liquid-vapor interface on the left of the plate. However, this is compensated by an equivalent decrease of liquid-vapor interface on the right of the plate. A non-vanishing effect is that the vertical displacement increases the length of dry plate by $dz/\cos \phi(0)$, while the wetted part is decreased by the same amount. Assuming there is no contact angle hysteresis ($\Delta \theta=0$), such that $\gamma_{SV}-\gamma_{SL}=\gamma \cos \theta_e$, one finds the forces due to the left and right side:
\begin{eqnarray}
F_z^L&=&\frac{\gamma (\sin \phi(0)-\cos \theta_e)}{\cos \phi(0)}, \\
F_z^R&=&\frac{\gamma (-\sin \phi(0)-\cos \theta_e)}{\cos \phi(0)}.
\end{eqnarray}
Here we introduced the shorthand $\gamma=\gamma_{LV}$, which will be employed in the remainder of the paper. Again, these forces are per unit width of the plate. Similarly, by moving the plate horizontally by $dx$, one reduces the water area by $\,dx$ so that:
\begin{equation}
F_x^L=- \gamma, \quad{\rm and}\quad F_x^R= \gamma.
\end{equation}
The total horizontal force $F_x$ thus vanishes while the total vertical force reads:
\begin{equation}\label{eq:Fz}
F_z = F_z^L + F_z^R =-\frac{2 \gamma \cos \theta_e}{\cos \phi(0)}.
\end{equation}
These results are easily generalized to incorporate contact angle hysteresis, i.e. allowing for different contact angles to the left and to the right of the plate~\cite{ParkJFM}. For this, one replaces $2\cos \theta_e$ by $\cos \theta_L + \cos \theta_R$ in equation (\ref{eq:Fz}), where $\theta_{L}$ and $\theta_{R}$ denote the angles on left and right. Note that $F_z$ can be also be obtained by considering the force contributions around each of the contact lines (cf. Sec.~\ref{sec:meniscus}). 

Combining (\ref{eq:el1}) with the $F_z$ induced by gravity and surface tension derived above, one obtains the equation for the plate
\begin{equation}\label{eq:el2}
B \phi''=\left(\rho gH s + 2\gamma \frac{\cos \theta_e}{\cos \phi(0)}\right) \sin \phi.
\end{equation}
To analyze this equation it is convenient to introduce a length scale expressing the strength of gravity with respect to the bending stiffness:
\begin{equation}
\ell_{\rm \chi} = \left( \frac{B}{\rho g H} \right)^{1/3}.
\end{equation}
The analysis implicitly assumes that $\ell_{\rm \chi}$ is much larger than $H$.

We show below that this is the typical scale over which the plate is curved. For the elastomeric plate used in the experiment presented in Fig.~\ref{fig:analysis}, $\ell_{\rm \chi}$ is around $7~$mm. Scaling the curvilinear coordinate as 
\begin{equation}
S=\frac{s}{\ell_\chi}, \quad {\rm and} \quad \Phi(S) = \phi(s),
\end{equation}
the torque balance (\ref{eq:el2}) becomes
\begin{equation}\label{eq:el3}
\Phi''=\left(S+ 2S_0 \frac{\cos \theta_e}{\cos \phi(0)}\right) \sin \Phi.
\end{equation}
Here we introduced a dimensionless number
\begin{equation}
S_0= \frac{\gamma \ell_\chi^2}{B} = \left( \frac{\ell_\chi}{\ell_{EC}}\right)^2,
\label{def_S0}
\end{equation}
The dimensionless number $S_0$ expressed in (\ref{def_S0}) manifests the importance of the vertical surface tension forces with respect to the bending stiffness. In the experiment presented in Fig.~\ref{fig:analysis}, $\ell_{EC}$ is around $7.5$~mm so that $S_0$ is of order unity  ($S_0\simeq0.90$). 

Interestingly, $S_0$ can be interpreted as the ratio of $\ell_\chi$, the ``effective" length of the plate, and the elastocapillary length $\ell_{EC}$. This is consistent with the general picture of elastocapillarity, namely that surface tension can induce deformations (such as buckling), when the plate is longer than $\ell_{EC}$~\cite{NeukJMPS,RomanJPCM,TawLANGMUIR,HazelPRSA,ChiodiEPL}. Below we will see that under partial wetting conditions, there is another dimensionless parameter associated to the torques induced by normal forces. 

Equation (\ref{eq:el3}) can be solved analytically when the angle of deflection is small, i.e. when $\sin\Phi(0) \simeq \Phi(0)$ and $\cos\Phi(0)\simeq1$. This limit is relevant at large distances from the meniscus, where the plate tends to a straight line (Fig.~\ref{fig:analysis}), as well as for describing the onset of the instability (Sec.~\ref{sec:instability}). The equation then becomes
\begin{equation}
 \Phi''=\left[S+2 S_0 \cos \theta_e \right] \Phi,
\end{equation}
which can be solved as:
\begin{equation}\label{eq:airy}
\Phi(S)=\Phi_1\;{\rm Ai}\left(S+2S_0 \cos \theta_e\right),
\end{equation}
where ${\rm Ai}$ is the Airy function. The integration constant $\Phi_1$ determines the amplitude of the deflection and has to be solved from the boundary condition at the meniscus. The second Airy function ${\rm Bi}(S)$ diverges for large arguments and thus does not comply with the boundary condition $\phi(\infty)=0$. Using the large $S$ asymptotics of ${\rm Ai}$, we find 
\begin{equation}
\Phi(S) \simeq \Phi_1 \frac{e^{-\frac23S^{3/2}}}{2\sqrt\pi\,S^{1/4}}.
\end{equation}
The plate thus naturally tends to a vertical line. Realizing that $S=s/\ell_\chi$, we indeed find that $\ell_\chi$ sets the length scale over which the deflection decays along the upward direction. 

\subsection{The meniscus region}\label{sec:meniscus}

The plate outside the liquid is described by a second order ordinary differential equation and thus requires two boundary conditions. A first boundary condition is $\phi(\infty)=0$, which, for example, was used while deriving (\ref{eq:airy}). The second boundary condition comes from the torques exerted at the meniscus region. As shown in (\ref{eq:Mi}), the internal torque experienced by the plate is proportional to the curvature $d\phi/ds$, which balances the external torque ${\cal T}_e$ applied by the meniscus:
\begin{equation}
{\cal T}_i + {\cal T}_e = 0.
\end{equation}
This boundary condition has to be evaluated at the position of the upper contact line. Namely, this point marks the edge of the domain for (\ref{eq:el1}), for which no normal forces were taken into account along the plate. In the paragraphs below we assume the meniscus on the left is higher than that on the right, as in Fig.~\ref{fig:analysis}. The left and right contact line positions, $z_L$ and $z_R$, are found from the classical meniscus solutions \cite{deGennes}
\begin{eqnarray}\label{eq:rise}
z_L &=& \pm \ell_\gamma \left[ 2(1-\sin(\theta_L - \phi(s_L)) \right]^{1/2}, \\
z_R &=& \pm \ell_\gamma \left[ 2(1-\sin(\theta_R + \phi(s_R)) \right]^{1/2}, 
\end{eqnarray}
where the sign (symbol $\pm$) depends on the value of the contact angle $\theta_e$ with respect to $\pi/2$ or equivalently on the sign of $\cos \theta_e$. Note that we now allow explicitly for different contact angles on both sides of the plate. The length scale of the meniscus is given by the capillary length
\begin{equation}
\ell_\gamma = \left( \frac{\gamma}{\rho g} \right)^{1/2}
\end{equation}
and reflects the balance between surface tension and the hydrostatic pressure (gravity). In the conditions of Fig.~\ref{fig:analysis}, $\ell_\gamma$ is around $2.3\,$mm. The analysis is simplified by using the hierarchy of length scales:
\begin{equation}
H \ll \ell_\gamma \ll \ell_\chi \sim \ell_{EC},
\end{equation}
Since $\ell_\gamma$ is significantly smaller than $\ell_\chi$, it is natural to use a different scaling for the meniscus region. To avoid confusion with the preceding paragraph, where we scaled the curvilinear coordinate $S=s/\ell_\chi$, we scale only the cartesian coordinate in the meniscus region:
\begin{equation}
Z= \frac{z}{\ell_\gamma}.
\end{equation}
One can assume $\ell_\gamma \ll \ell_{EC}$, which suggests that the length scale over which the capillary forces are assumed to be influential is substantially smaller than the length scale over which the capillary-force induced bending can be significant. As a consequence, one can assume that the plate represents a negligible curvature in the meniscus region, so that the angle can be considered constant, $\phi(s_L)=\phi(s_R)=\phi(0)$. This gives a simple relation between the coordinates $z=s \cos \phi(0)$. 

\subsubsection{Capillary forces}

Before addressing the torques, we first specify the various capillary forces exerted by the liquid on the solid plate. The detailed spatial distribution of capillary forces is a difficult question in itself, as addressed e.g. in~\cite{DasPOF,MarchandAJP}. As will be commented in Sec.~\ref{sec:distribution}, the resultant forces can be represented as shown in the inset of Fig.~\ref{fig:analysis}. First, there is a force per length of magnitude $\gamma$ that pulls along the liquid-vapor interface. Second, there is a contribution due to hydrostatic pressure in the liquid, which is unbalanced whenever the two contact lines are at a different height (i.e. when $z_L\neq z_R$). This pressure is acting normal to the solid surface and has to be integrated between the two contact lines.

Projecting the tangential force contributions, one finds the resultant force along the plate
\begin{equation}\label{eq:Fs}
F_s = - \gamma \left( \cos \theta_L + \cos\theta_R \right), 
\end{equation}
taken in the positive $s$ direction. For $\cos \theta_e < 0$, equilibrium angle $\theta_e > \pi/2$, this force is compressing the plate. Similar to the classical buckling instability, such a compressive force has a destabilizing effect. For $\cos \theta_e > 0$ it is stabilizing. 

The normal forces add up to 
\begin{equation}\label{eq:bla}
F_n = \gamma \left( \sin \theta_R - \sin \theta_L \right) + F_p, 
\end{equation}
where $F_p$ is the unbalanced hydrostatic pressure appearing on the left of the plate. Assuming that $z_L > z_R$, or equivalently $\theta_L < \theta_R$, this hydrostatic pressure is obtained by integration as 
\begin{eqnarray}\label{eq:Fp}
F_p &=& \int_{s_R}^{s_L} ds \, p(s) \nonumber \\
&=& - \int_{s_R}^{s_L} ds \, \rho g s \cos\phi(0) \nonumber \\
&=& - \frac{1}{2}  \rho g \left(s_L^2 - s_R^2 \right) \cos \phi(0) \nonumber \\
&=& - \frac{1}{2}  \gamma \frac{Z_L^2 - Z_R^2}{\cos \phi(0)},
\end{eqnarray}
where we in the last step we used $\rho g = \gamma/\ell_\gamma^2$ and $s = Z \ell_\gamma /\cos \phi(0)$. These equation can be further worked out using (\ref{eq:rise}), where we take $\phi(s_L)=\phi(s_R)=\phi(0)$. Combined with (\ref{eq:bla}) this finally gives
\begin{equation}\label{eq:Fn}
F_n = -  \gamma \left( \cos \theta_L + \cos\theta_R \right) \tan \phi(0). 
\end{equation}

Let us emphasize that these resultant force components (\ref{eq:Fs},\ref{eq:Fn}) can be projected in the $(x,z)$ directions in order to compare to the virtual work result discussed in Sec.~\ref{sec:outside}. Indeed, the projection gives $F_x=0$ while one recovers the correct $F_z$ upon replacing $2\cos \theta_e$ by $\cos \theta_L + \cos \theta_R$ in (\ref{eq:Fz}). This illustrates the importance of the force due to the unbalanced hydrostatic pressure $F_p$. Its magnitude is of order $\gamma$ [see (\ref{eq:Fp})] and involves the expressions that depend on the contact angles $\theta_L$ and $\theta_R$. Most importantly, only by adding $F_p$ to $\gamma(\sin{\theta_R}-\sin{\theta_L})$, one recovers the capillary forces obtained from the virtual work principle [see section \ref{sec:outside}]. Therefore $F_p$ should be interpreted as a capillary force.

\subsubsection{Capillary torques}

Having established the capillary forces in the meniscus region, we are in a position to compute the associated torques. Since we are interested in the boundary condition for the plate outside the reservoir, we compute the torque around the highest contact line, i.e. $s_L$. Using the convention that positive torques induce a rotation in clockwise direction, the normal forces then give a torque

\begin{eqnarray}
{\cal T}_n &=& -   \gamma (s_L - s_R) \sin \theta_R  \nonumber \\
 &=& -  \gamma \ell_\gamma \frac{(Z_L -Z_R) \sin \theta_R}{\cos \phi(0)}.
\end{eqnarray}
From the construction in the inset of Fig.~\ref{fig:analysis} it is clear that this torque is destabilizing. Namely, if we consider a small perturbation where the plate is slightly bent to the right, the meniscus on the left rises higher than the meniscus on the right. As a consequence the surface tension force on the right has a larger moment arm than that its counterpart on the left. The induced torque on the plate acts in the same direction as the initial perturbution, and hence, has a destabilizing effect.

Similarly to (\ref{eq:Fp}), there is a torque induced by the hydrostatic pressure. This is obtained by integrating over the pressure, now including a moment arm $s_L - s$:
\begin{eqnarray}\label{eq:Mp}
{\cal T}_p &=& \int_{s_R}^{s_L} ds \, p(s) (s_L - s)\nonumber \\
&=& \int_{s_R}^{s_L} ds \, \rho g s \cos\phi(0) (s - s_L)\nonumber \\
&=& \frac{1}{6} \gamma \ell_\gamma \frac{(Z_L + 2Z_R)(Z_L-Z_R)^2}{\cos^2 \phi(0)}.
\end{eqnarray}
Interestingly, this torque scales as $(Z_L-Z_R)^2$, which reflects the fact that both the integrated pressure and the moment arm are proportional to $Z_L-Z_R$. For small asymmetry we can thus neglect ${\cal T}_p$ with respect to the moment induced by the force at the contact line. 

Finally, the torque induced by the tangential forces is strictly zero when $\Delta \theta=0$, as the forces act in the same directions. For small hysteresis, the resultant torque is of order $\sim \gamma H \Delta \theta$, since the arm for the tangential force is half the thickness of the plate. Clearly, this can be neglected with respect to ${\cal T}_n$, for which the arm is given by $\ell_\gamma$. To summarize, we find the external torque 
\begin{eqnarray}
{\cal T}_e &=& {\cal T}_n + {\cal T}_p,
\end{eqnarray}
which for small $\phi(0)$ and small hysteresis is dominated by ${\cal T}_n$.

\subsection{Dimensionless equations}\label{sec:dimensionless}

The results of the preceding paragraphs can be summarized as follows. We found that the plate in the region outside the bath is governed by the length scale $\ell_\chi$, which in practice is much larger than the size of the meniscus $\ell_\gamma$. To separate the regimes, we use the dimensionless curvilinear coordinate $S=s/\ell_\chi$ outside the bath, for which the shape can be solved from (\ref{eq:el3}), i.e. 

\begin{equation}\label{eq:el4}
\Phi''=\left(S+ 2S_0 \frac{\cos \theta_e}{\cos \Phi(0)}\right) \sin \Phi.
\end{equation}
This it to be complemented by a boundary condition at $S=s_L/\ell_\chi \approx 0$, since $s_L$ is of the order of $\ell_\gamma \ll \ell_\chi$. This boundary condition is most conveniently expressed in terms of $Z=z/\ell_\gamma$. When $Z_L > Z_R$ this gives 
\begin{eqnarray}\label{eq:bc}
\Phi'(0)= \nonumber \\
T_0\;\left(-\frac{(Z_L-Z_R) \sin \theta_R}{\cos \Phi(0)}+ 
\frac{(Z_L+2 Z_R)(Z_L-Z_R)^2}{6\cos^2 \Phi(0)} \right), \nonumber \\
&&
\end{eqnarray}
while for $Z_R > Z_L$ one has
\begin{eqnarray}
\Phi'(0) = \nonumber \\
T_0\;\left(\frac{(Z_L-Z_R) \sin \theta_L}{\cos \Phi(0)} - 
\frac{(2Z_L+Z_R)(Z_L-Z_R)^2}{6\cos^2 \Phi(0)} \right), \nonumber \\
&&
\end{eqnarray}
where
\begin{equation}
T_0= \frac{\ell_\chi \ell_\gamma}{\ell_{EC}^2}.
\end{equation}
In the experiment presented in Fig.~\ref{fig:analysis}, $T_0$ is of order unity ($T_0\simeq0.285$). The values of $Z_{L,R}$ are determined by the contact angles from (\ref{eq:rise}), and the plate inclination at the bottom $\Phi(0)$. The latter parameter follows as a result of the calculation and can be used to identify the buckling instability. 

Apart from the contact angle $\theta_e$ (and the hysteresis $\Delta \theta$), the problem is governed by two dimensionless parameters that can be interpreted as a ratio of length scales:
\begin{equation}
S_0= \left( \frac{\ell_\chi}{\ell_{EC}} \right)^2, \quad \quad 
T_0= \frac{\ell_\chi \ell_\gamma}{\ell_{EC}^2}.
\end{equation}
The first of these parameters can be interpreted as the ability to induce \emph{buckling} for the tangential capillary force, provided that $\cos \theta <0 $. Consistent with the standard view of elastocapillarity, this effect is governed by the elastocapillary length $\ell_{EC}$ with respect to the ``effective" length of the plate, set by $\ell_\chi$. The second dimensionless parameter sets the strength of the \emph{bending} induced by the torque generated in the meniscus. As it involves a torque, the capillary length $\ell_\gamma$ intervenes as the moment arm.

\section{Bifurcations}\label{sec:instability}

It is clear that the straight plate, $\Phi(S)=0$, is a solution of (\ref{eq:el4}) that satisfies the boundary condition (\ref{eq:bc}). We now analyze the stability of these solutions in terms of the parameters $S_0$ and $T_0$, for different values of $\theta_e$. Throughout this section we assume no hysteresis, i.e. $\Delta \theta=0$ or equivalently $\theta_L = \theta_R = \theta_e$. We first perform a linear analysis to identify the threshold and discuss the regimes where bending or buckling are dominant. Subsequently, we numerically compute the bifurcation diagram by following the various solution branches in the nonlinear regime.

 \subsection{Instability threshold}

The threshold of instability of the straight plate is obtained by linearizing the problem for small $\Phi$, as already done in (\ref{eq:el3}), yielding a solution
\begin{equation}\label{eq:airy2}
\Phi(S) = \Phi_1 {\rm Ai}(S+2S_0 \cos \theta_e).
\end{equation}
Similarly, the boundary condition (\ref{eq:bc}) can be expanded as, 
\begin{equation}\label{eq:bc2}
\Phi' (0) \simeq -2 T_0 \frac{|\cos \theta_e| \sin \theta_e}{\sqrt{2(1-\sin \theta_e)}}  \Phi(0),
\end{equation}
where the expression for the meniscus rise (\ref{eq:rise}) was used. Combining (\ref{eq:airy2}) and (\ref{eq:bc2}) one obtains the equation for a ``neutral mode", which is a solution of the deflection profile for arbitrary (small) perturbation amplitude $\Phi_1$:
\begin{equation}\label{eq:onset}
2T_0\frac{\sin \theta_e |\cos \theta_e|}{\sqrt{2(1-\sin \theta_e)}}+ \frac{{\rm Ai'}\left[2 S_0 \cos \theta_e\right]}{{\rm Ai}\left[2 S_0 \cos \theta_e\right]}=0.
\end{equation}

Indeed, this equation provides the threshold for the instability in terms of the parameters $T_0$, $S_0$ and $\theta_e$. This can be seen, e.g. by varying one of these parameters while keeping the other two constant. One finds that the internal moment ${\cal T}_i$ dominates the external torque ${\cal T}_e$ (stable) or vice versa (unstable), as the parameter is varied across the neutral condition (\ref{eq:onset}). 

\subsection{Bending instability: $\theta_{e} < \pi/2$}

We now reveal the destabilizing effect of the torque in the meniscus, associated to the parameter $T_0$, which tries to bend the plate. This mechanism is most relevant for $\theta < \pi/2$, for which it turns out the \emph{only} destabilizing mechanism: the buckling parameter $S_0$ is stabilizing in this range as it multiplies with $\cos \theta_e > 0$. In this context of bending, (\ref{eq:onset}) indeed provides the critical $T_0$ beyond which the flat solution becomes unstable. Namely, for larger $T_0$ the torque in the meniscus ${\cal T}_n$ becomes larger than the internal torque of the plate ${\cal T}_i$ for small perturbations, hence leading to instability. 
\begin{figure}[t]
\begin{center}
\includegraphics{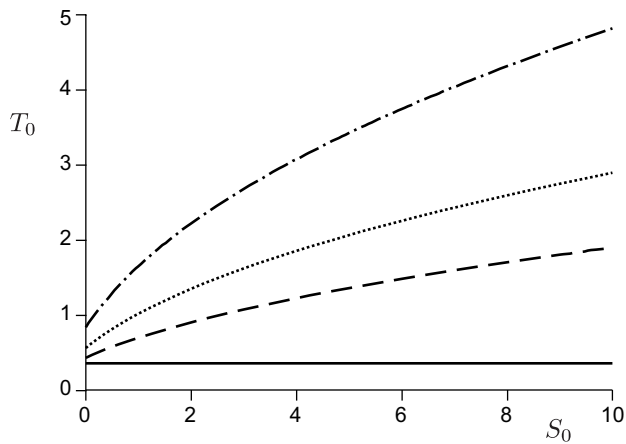}
\caption{\label{fig:bendingonset} The threshold of stability $T_0$ versus $S_0$ for different values of the contact angle $\theta_e = \pi/6, \pi/4,\pi/3, \pi/2$. For $\cos \theta_e >0$, the bending threshold $T_0$ increases with $S_0$ since the vertical capillary force has a stabilizing effect. The limiting case $\cos \theta_e=0$ has $F_z=0$, for which the bending threshold is independent of $S_0$.}
\end{center}
\end{figure}

The result of the stability analysis is shown in Fig.~\ref{fig:bendingonset}, depicting the critical $T_0$ versus $S_0$ for several contact angles below $\pi/2$. One observes the following trends. First, upon increasing $\theta_e$ the instability is triggered at a smaller $T_0$. This occurs as the destabilizing normal forces are proportional to $\sin \theta_{e}$, and thus becomes more influential for larger contact angles. Second, the instability threshold increases with $S_0$, which represents the strength of the tangential capillary forces. In the regime $\theta_e<\pi/2$ or $\cos \theta_e >0$, these tangential forces are pulling on the plate and are indeed stabilizing; hence, one requires a larger value for $T_0$ to induce the instability. In the limit of large $S_0$ one can expand the Airy functions, yielding the asymptotics $T_0 \sim S_0^{1/2}$. Finally, for the limiting case where the tangential capillary forces vanish, $\theta_e=\pi/2$, the critical $T_0$ does not depend on $S_0$ and can be computed analytically as
\begin{equation}\label{eq:derrida}
T_0(\pi/2) = - \frac{{\rm Ai}'(0)}{2{\rm Ai}(0)}= \frac{3^{1/3} \Gamma(2/3)}{2\Gamma(1/3)} \approx 0.3645 \cdots 
\end{equation}

\begin{figure}[t]
\begin{center}
\includegraphics{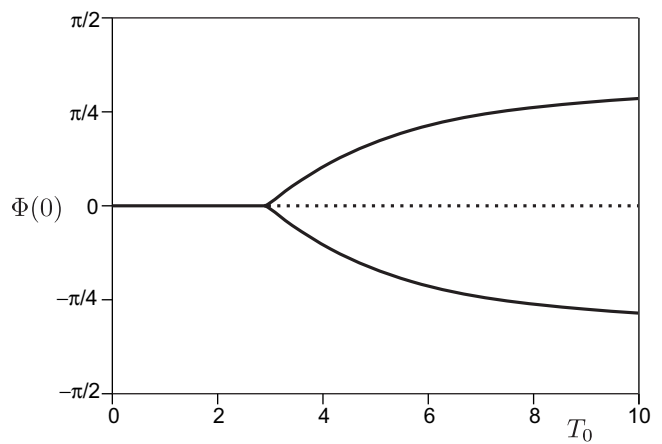}
\caption{\label{fig:bifurcationbending} Typical bifurcation diagram for bending induced instability, obtained from numerical integration of (\ref{eq:el4},\ref{eq:bc}). Solutions branches characterized by the angle at the bottom of the plate, $\Phi(0)$, upon varying the bending strength $T_0$. The values of $S_0=10.0$ and $\theta_e=\pi/4$ were kept fixed.}
\end{center}
\end{figure}

We now analyze the non-linear behavior of the solutions above the threshold. We numerically solve (\ref{eq:el4},\ref{eq:bc}) and characterize the various solutions with $\Phi(0)$, the plate inclination at the bottom. Figure~\ref{fig:bifurcationbending} shows a typical bifurcation diagram for the bending induced instability, by depicting the variation of $\Phi(0)$ with $T_0$. One recognises a supercritical pitchfork bifurcation, with a critical exponent $1$. The diagram is for a given value of $S_0$ and $\theta_e$, and shows that there is a critical $T_0$ beyond which the trivial solution $\Phi(0)=0$ becomes unstable and bifurcates into two stable branches. Above threshold, the plate inclination saturates at a finite angle $\Phi(0)$. The critical $T_0,$ can be easily read off from Fig.~\ref{fig:bendingonset} by drawing a line parallel to $T_0$ axis, passing through the corresponding $S_0$ (here $S_0=10$) and obtaining the $T_0$ from the point of intersection of this line (here $\theta_e=\pi/4$).

\subsection{Buckling instability: $\theta_{e} > \pi/2$}\label{sec:buckling}

We now consider the case $\theta_{e} > \pi/2$ for which the tangential forces are compressing the plate and can lead to the classical buckling instability. To isolate this buckling from the bending we can consider the case where $T_0 \ll1$ or $\theta_e \approx \pi$. According to (\ref{eq:onset}), the onset of buckling is associated to the rightmost maximum of the Airy function, i.e. ${\rm Ai}'(c_0) =0$, which gives 
\begin{figure}[t]
\begin{center}
\includegraphics{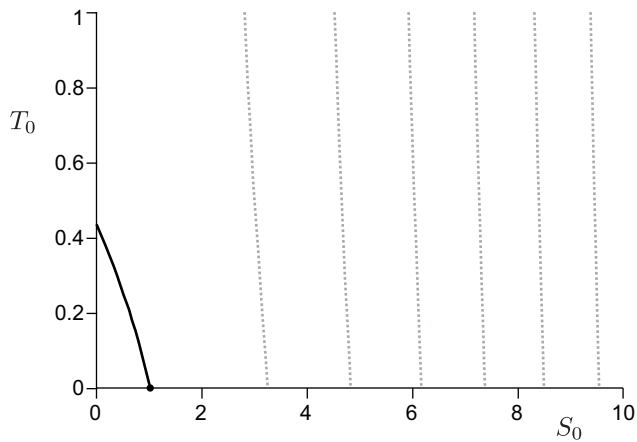}
\caption{\label{fig:bucklingonset} The threshold of stability $T_0$ versus $S_0$ for $\theta_e = 2\pi/3$. The vertical force $F_z$ now compresses the wire and leads to buckling, even in the absence of bending ($T_0=0$). Similar to the classical buckling instability, the higher order branches correspond to all extrema of the Airy function.}
\end{center}
\end{figure}
\begin{equation}
S_0 =\frac{c_0}{2\cos \theta_e}, \quad {\rm with} \quad c_0 = -1.01886 \cdots.
\end{equation}
This value is indicated by the closed circle in Fig.~\ref{fig:bucklingonset}. It can be seen in the figure that for $T_0\neq 0$, the threshold is lowered due to the destabilizing nature of the torque in the meniscus region.

Beyond the onset, one observes a sequence of branches associated to the other maximima and minima of the Airy function, at more negative arguments. In analogy to the classical buckling, these correspond to the higher order modes. The stability threshold is obtained from (\ref{eq:onset}) which gives, for $T_0\ll1$, ${\rm Ai}'(c_n)=0$, so that 
\begin{equation}
S_0 =\frac{c_n}{2\cos \theta_e}.
\end{equation}
The higher order branches correspond to large arguments of the Airy function, and can be determined accurately from asymptotics of ${\rm Ai}(s)$:
\begin{equation}
c_n \simeq - \left[ \frac{3\pi}{8} (4n+1)    \right]^{2/3}.
\end{equation}
%
%When $n$ becomes odd the threshold changes from unstable to stable, whereas when $n$ becomes even the threshold changes from stable to unstable. 
%
\begin{figure}[t]
\begin{center}
\includegraphics{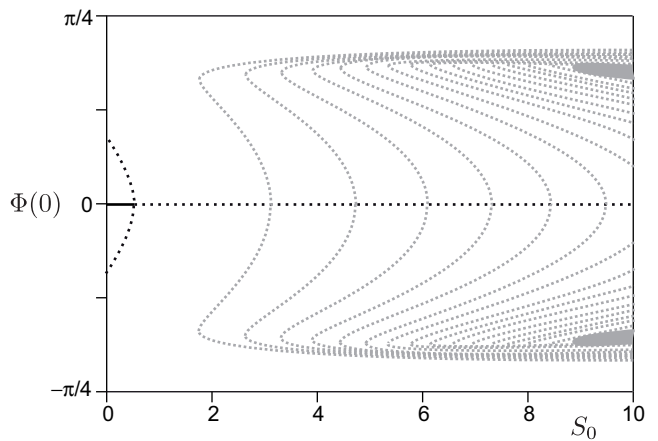}
\caption{\label{fig:bifurcationbuckling} Typical bifurcation diagram for the buckling induced instability, obtained from numerical integration of (\ref{eq:el4},\ref{eq:bc}). Solutions branches characterized by the angle at the bottom of the plate, $\phi(0)$, upon varying the buckling strengths $S_0$. The values of $T_0=0.25$ and $\theta_e=2\pi/3$ were kept fixed.}
\end{center}
\end{figure}

\begin{figure}[t]
\begin{center}
\includegraphics{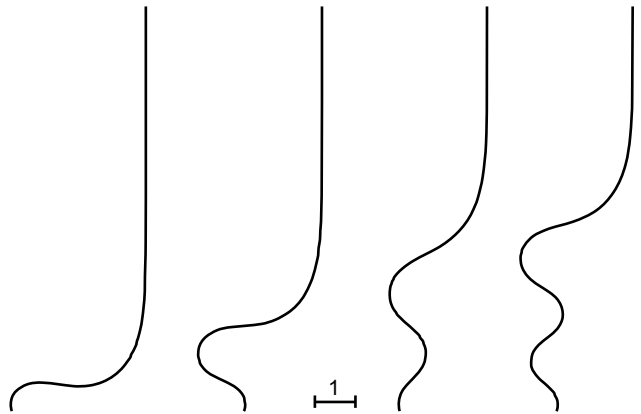}
\caption{\label{fig:shapes} Shapes of the plate with $\Phi(0)=0$ for  different buckling strengths $S_0$. From left to right: $S_0=2.689$, $S_0=4.302$, $S_0=5.636$, $S_0=6.830$. The values of $T_0=0.25$ and $\theta_e=2\pi/3$ are as in Fig.~\ref{fig:bifurcationbuckling} and were kept fixed.}
\end{center}
\end{figure}

Similar to Fig.~\ref{fig:bifurcationbending}, in Fig.~\ref{fig:bifurcationbuckling}, we illustrate a typical bifurcation diagram for the buckling induced ($\theta_e>\pi/2$) instability and characterize the various solution with $\Phi(0)$. For the first branch, as we move along $\Phi(0)=0$, on crossing a critical $S_0$, the solution becomes unstable and produces two additional unstable solutions. This implies that there is no stable solution at a finite angle. Physically, this suggests that on slight perturbation from its equilibrium position, the wire (or plate) ends up on the surface [as in Fig.~\ref{fig:bending}], corresponding to $\Phi(0)=\pi/2$. On moving further along $\Phi(0)=0$, one encounters further bifurcations which correspond to higher order modes becoming unstable. Similarly to the previous bifurcation diagram, the critical values for $S_0$ are obtained from Fig.~\ref{fig:bucklingonset}, by drawing a line parallel to $S_0$ axis, passing through $T_0=0.25$. Typical solutions with $\Phi(0)=0$ are depicted in Fig.~\ref{fig:shapes}. As in classical buckling, the successive branches are separated by half a wavelength.

\section{Experimental perspective}\label{sec:exp}

\subsection{Influence of thickness}
In this section we would like translate the analysis in terms of dimensionless numbers $S_0$ and $T_0$ to an experimental situation. As an illustration, we consider a case where we fix the material properties of the liquid and the elastic solid, and vary the thickness $H$. The instability is then reached below a critical thickness, for which the bending rigidity is sufficiently weak. Alternatively, one may perform an experiment as sketched in Fig.~\ref{fig:bending}, where the thickness is fixed but the contact angle is varying. 

Retracing the steps of the analysis, one finds that $S_0$ and $T_0$ are constructed from the thickness of the plate $H$ and the material parameters $\gamma$, $E$, $(\rho g)$. The bending stiffness $B$ by itself is not a material parameter as it depends on the thickness as $B=EH^3/12$. By selecting the properties of the liquid and the elastic solid, one fixes two length scales
\begin{equation}
\ell_{EG} = \frac{E}{12\rho g}, \quad {\rm and} \quad 
\ell_\gamma= \left( \frac{\gamma}{\rho g} \right)^{1/2}.
\label{length_scales}
\end{equation}
In the experiment presented in Fig.~\ref{fig:analysis}, $\ell_{EG}$ is around $56$~cm and the capillary length is $\ell_\gamma=2.3$mm so that $\ell_{EG}/\ell_{\gamma}\simeq250$. We can express $T_0$ and $S_0$ in terms of the length scales defined in (\ref{length_scales}) as:
\begin{equation}
T_{0}=\frac{(H^2\ell_{EG})^{1/3}{\ell_{\gamma}}^3}{\,\ell_{EG}H^3}=H^{-7/3}\,{\ell_{EG}}^{-2/3}\,{\ell_{\gamma}}^3,
\label{T0_ln_sc}
\end{equation}
\begin{equation}
S_{0}=\frac{(H^2\ell_{EG})^{2/3}{\ell_{\gamma}}^2}{\,\ell_{EG}H^3}=H^{-5/3}\,{\ell_{EG}}^{-1/3}\,{\ell_{\gamma}}^2.
\label{S0_ln_sc}
\end{equation}
Interestingly, the length scales $\ell_{EG}$ and $\ell_\gamma$ appear in different combinations in the two parameters. We thus anticipate that the threshold thickness $H$  presents different  scaling laws, depending on whether the instability is due to bending ($T_0$) or due to buckling ($S_0$).
\begin{figure}[t]
\begin{center}
\includegraphics{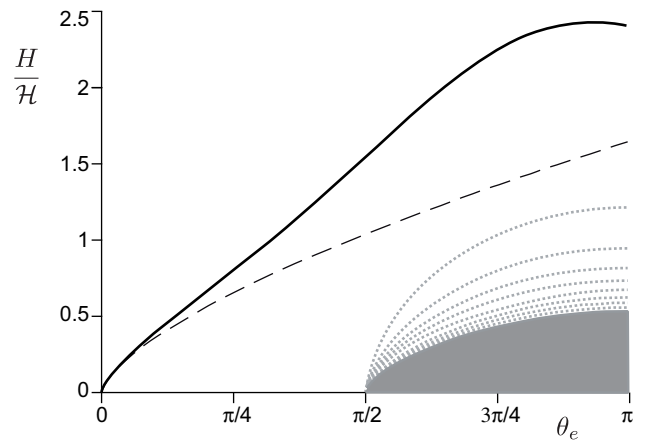}
\caption{\label{fig:h} Experimental perspective: the threshold of instability by varying the plate thickness $H$ and the contact angle $\theta_e$, for $\ell_{EG}/\ell_{\gamma}=250$. The plate thickness $H$ is normalized by ${\cal H}$, defined in (\ref{eq:calh}) as a threshold value for intermediate $\theta_e$. For typical experimental conditions, ${\cal H}=0.47$~mm. }
\end{center}
\end{figure}

In Fig.~\ref{fig:h} we show the critical thickness for different values of the contact angle. We scaled the thickness by assuming $T_0 \sim {\cal O}(1)$, which implies a characteristic thickness ${\cal H}$:
\begin{equation}\label{eq:calh}
{\cal H} =  \ell_\gamma^{9/7}  \ell_{EG}^{-2/7}.
\end{equation}
In the experiment presented in Fig.~\ref{fig:analysis}, ${\cal H}$ is around $0.47$~mm. The upper line in Fig.~\ref{fig:h} represents the threshold thickness for instability. The main trend of the graph is that the instability is more difficult to reach, i.e. requires a smaller thickness of the plate, as the contact angle $\theta_e$ is decreased. 

A key result, however, is that even for very small contact angles there is still an instability due to the torque exerted by the normal forces. This contrasts the classical buckling picture, since for small $\theta_e$ the capillary forces are not compressing, but are in fact pulling on the plate. Note that in this regime of small $\theta_e$, $T_0\rightarrow\infty$, and therefore $\cal H$, as defined in (\ref{eq:calh}), is no longer the correct length scale for the threshold thickness. Rather, in this regime the critical thickness is obtained from,
\begin{eqnarray}
&&2\left(\frac{\mathcal H}{H}\right)^{7/3}\frac{\sin \theta_e |\cos \theta_e|}{\sqrt{2(1-\sin \theta_e})}+ \nonumber \\
&&\frac{{\rm Ai'}\left[2\left(\frac{\mathcal H}{H}\right)^{5/3}\left(\frac{\ell_{EG}}{\ell_{\gamma}}\right)^{1/7} \cos \theta_e\right]}{\rm Ai \left[2\left(\frac{\mathcal H}{H}\right)^{5/3}\left(\frac{\ell_{EG}}{\ell_{\gamma}}\right)^{1/7} \cos \theta_e\right]}=0\;.
\end{eqnarray}
For $\theta_e\rightarrow0$ and $H/\mathcal{H}\rightarrow0$:
\begin{align}
\sqrt{2}\left(\frac{\mathcal H}{H}\right)^{7/3} \theta_e - \left[2\left(\frac{\mathcal H}{H}\right)^{5/3}\left(\frac{\ell_{EG}}{\ell_{\gamma}}\right)^{1/7}\right]^{1/2}=0\;,
\end{align}
which yields:
\begin{align}\label{eq:small}
\frac{H}{\mathcal H}=\left(\frac{\ell_{\gamma}}{\ell_{EG}}\right)^{1/21} {\theta_{e}}^{2/3}\;.
\end{align}
This asymptotic form is shown as the dashed line in Fig.~\ref{fig:h}. Indeed, this regime involves a different combination of $\ell_{EG}$ and $\ell_\gamma$ than ${\cal H}$. 

Above $\pi/2$, one enters the usual buckling regime. The graph also reveals the higher order buckling modes. From the previous paragraphs, a good approximation can be obtained as:
\begin{equation}\label{eq:buczone}
\frac{H}{\mathcal H}=\left(\frac{\ell_{EG}}{\ell_{\gamma}}\right)^{1/21} \left[ \frac{3\pi}{8} (4n+1)\right]^{-2/5} \left(-2\cos \theta_{e}\right)^{3/35}\;.
\end{equation}

The onset is generated by the highest of the lines. Finally, note that there is an optimal contact angle, slightly before $\theta_{e}=\pi$, for which the instability is most easily reached. The scaling near $\theta_e=\pi$ turns out
\begin{equation}\label{eq:large}
\frac{H}{\mathcal{H}}=\left(\frac{2}{|c_0|}\right)^{3/5}\left(\frac{\ell_{EG}}{\ell_{\gamma}}\right)^{3/35}\approx1.4988\left(\frac{\ell_{EG}}{\ell_{\gamma}}\right)^{3/35}.
\end{equation}

To summarize, the critical thickness below which the plate becomes unstable increases with the contact angle $\theta_e$, except very close to $\theta_e=\pi$, where the thickness displays a maximum. Intriguingly, the dependence of the characteristic thickness on the material parameters is not universal, but depends on the contact angle. Three regimes can be identified, involving different combinations of $\rho g$, $\gamma$ and $E$. At very small contact angles, one finds from (\ref{eq:small}):
\begin{equation}
H\propto (\rho g)^{-1/3}\gamma^{2/3}E^{-1/3}{\theta_{e}}^{2/3},
\end{equation}
Close to $\theta_e=\pi$ one has (\ref{eq:large}):
\begin{equation}
%H \propto (\rho g)^{-3/5}\gamma^{2/5}E^{1/5},
H \propto (\rho g)^{-2/5}\gamma^{3/5}E^{-1/5},
\end{equation}
while at intermediate contact angles one has (\ref{eq:calh})
\begin{equation} 
H \propto (\rho g)^{-5/14}\gamma^{9/14}E^{-2/7}. 
\end{equation} 
\begin{figure}[t]
\begin{center}
\includegraphics{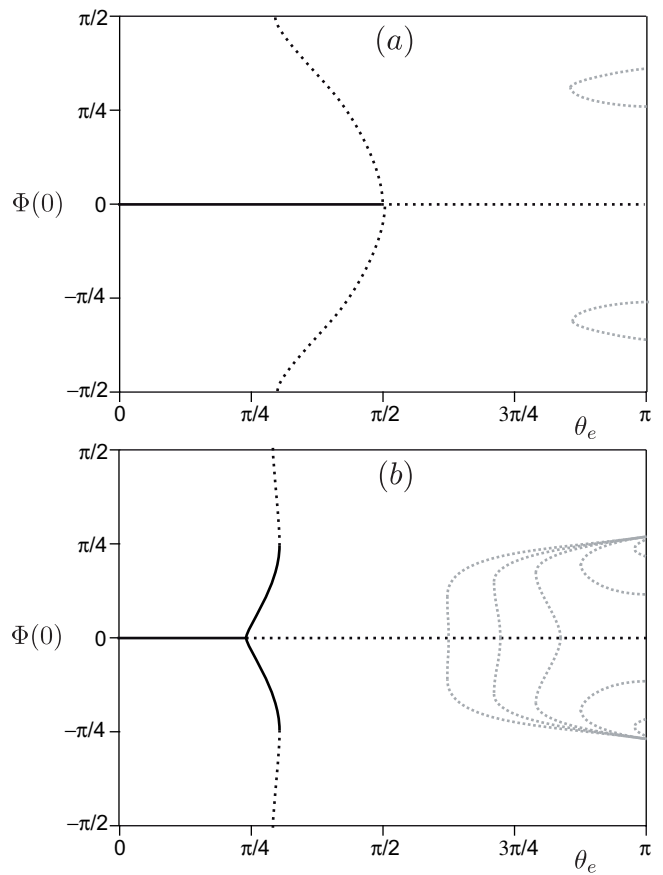}
\caption{\label{fig:h3} Bifurcation diagram, $\Phi(0)$ versus $\theta_e$ for $\ell_{EG}/\ell_{\gamma}=250$. The two panels correspond to different cross-sections of Fig.~\ref{fig:h}, namely (a) $H/{\cal H}=1.54$, and (b) $H/{\cal H}=0.77$.}
\end{center}
\end{figure}

\subsection{Influence of contact angle hysteresis}

Another important experimental feature is that one cannot eliminate a substantial hysteresis of the contact angle. This means that $\theta_L \neq \theta_R$, with typical experimental values $\Delta \theta \approx 0.1$ (in radians). Taking this hysteresis into account in the model, we find a small shift in the threshold of instability. However, the fact that hysteresis breaks the left-right symmetry of the problem has a much more pronounced effect on the general structure of the solutions and their bifurcation diagrams. 

To illustrate this, we first show two bifurcation diagrams corresponding to Fig.~\ref{fig:h} in the case without hysteresis ($\Delta \theta=0$). Figures~\ref{fig:h3}a and \ref{fig:h3}b are both obtained by varying $\theta_e$, for two different values of the plate thickness. The upper plot corresponds to $H/{\cal H}=1.54$, for which a single bifurcation is observed. The lower plot corresponds to $H/{\cal H}=0.77$, for which the higher order buckling modes are crossed. 

\begin{figure}[t]
\begin{center}
\includegraphics{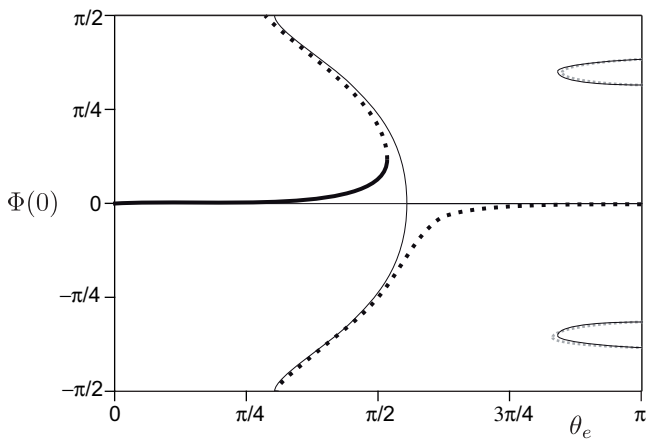}
\caption{\label{fig:h4} Effect on hysteresis on the bifurcation diagram, $\Phi(0)$ versus $\theta_e$, in the case $H/{\cal H}=1.71$ and $\ell_{EG}/\ell_{\gamma}=250$. $\Delta \theta = 0.1$ (thick lines) is compared to the case without hysteresis (thin line). The hysteresis allows for stable solutions of nonzero amplitude $\Phi(0)$.}
\end{center}
\end{figure}

The effect of contact angle hysteresis is revealed in Fig.~\ref{fig:h4}, comparing the bifurcation diagrams for $\Delta \theta=0$ and $\Delta \theta =0.1$. Clearly, the left-right symmetry of the problem is broken by the hysteresis, as reflected by the splitting of the branches for $\Phi(0)>0$ and for $\Phi(0)<0$. A practical consequence is that, below threshold, the stable solutions actually correspond to a nonzero $\Phi(0)$. This means that the stable states do not correspond to a straight plate: the plate is always slightly bent, even before the instability occurs. This effect is clearly visible in Fig.~\ref{fig:analysis}, which corresponds to a photograph of a stable plate that is indeed deformed towards a small angle. 

\section{Discussion}\label{sec:discussion}

\subsection{Spatial distribution of capillary forces}\label{sec:distribution}
The total force exerted by the liquid on the solid is $\gamma \cos \theta_e$ per unit contact line, as follows from a thermodynamic argument based on virtual work (Sec.~\ref{sec:formulation}). In this paper, we have treated these capillary forces as if they were perfectly localized at the contact line -- see the inset of Fig.~\ref{fig:analysis}. However, this is not necessarily a representation of the real distribution of capillary forces, since the solid can be submitted to a Laplace pressure wherever the solid surface is curved \cite{DasPOF,JeriPRL,MoraPRL}. In the case of the plate, for example, this curvature is localized at the bottom edges and could induce an upward force due to a Laplace pressure on the solid. To restore the thermodynamic resultant force, this necessarily means that additional downward forces must be present at the contact line to counteract this effect. Here we will not discuss the origin and nature of this Laplace pressure and we refer the interested reader to \cite{DasPOF,MoraPRL}~; assuming that it exists, how are the results derived in this paper affected? 

\begin{figure}[t]
\begin{center}
\includegraphics{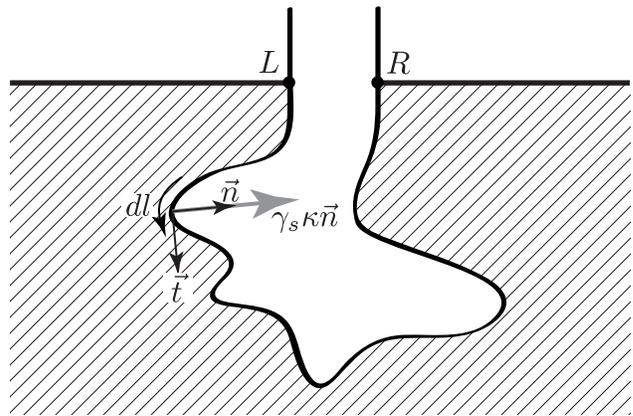}
\caption{\label{fig:SolidSurfaceTension} Sketch of the quantities required to compute the Laplace pressure exerted on a solid of arbitrary shape. See text for details.}
\end{center}
\end{figure}

We consider the two-dimensional situation depicted in Fig.~\ref{fig:SolidSurfaceTension}, where the bottom of the solid can take an arbitrary shape. The immersed part of the solid is submitted to a distribution of pressure $\gamma_s \kappa$ proportional to the curvature $\kappa$ and to the relevant surface tension coefficient $\gamma_s$. The total force exerted on the solid due to this curvature effect is written as a contour integral:
\begin{equation}\label{eq:blaaa}
\vec F_\kappa= \int_L^R \gamma_s \kappa \vec n\,dl=  \int_L^R \gamma_s \frac{d \vec t}{dl}\,dl=\left[\gamma_s \vec t\right]_L^R
\end{equation} 
where $l$ is the curvilinear coordinate, $\vec t$ is the local tangent vector and $\vec n$ the local normal vector. Elementary geometry gives $d\vec t/dl = \kappa \vec n$, where $\kappa$ is the local curvature of the surface. Hence, (\ref{eq:blaaa}) shows that the Laplace contribution to the total capillary force depends only on the tangent vectors at the contact line and is independent of the shape of the immersed solid. As mentioned above, this must be compensated by an additional force at the contact line to restore the thermodynamic result. Similarly, the total moment of the Laplace force  can be written  as a contour integral:
\begin{equation} 
\vec \tau_\kappa= \int_L^R \gamma_s \vec r \wedge \kappa \vec n\,dl=  \int_L^R \gamma_s   \frac{d \vec r \wedge \vec t}{dl}\,dl=\left[ \gamma_s  \vec r \wedge \vec t\right]_L^R
\end{equation} 
where we have used the property $d \vec r / dl=\vec t$. As for the resulting force, the moment also does not depend on the shape of the object and is equal to the moment of a force $\gamma_s \vec t$ that would be localized at the contact line. In other words, neither the total force nor the total moment exerted on the upper part of the plate depend on the distribution of capillary forces below the surface. 

We thus conclude that all results presented in this paper are perfectly insensitive to the true spatial distribution of the capillary forces. Reciprocally, it also implies that the Laplace pressure on a solid cannot be characterised using bending or buckling experiments. Instead, one must measure the deformations of the solid surface \cite{JeriPRL}. 

\subsection{Conclusion}

In this paper we identified two separate mechanisms that can lead to elastocapillary instability of a flexible plate partially immersed in a liquid. The tangential components of the capillary forces can induce buckling whenever the contact angle $\theta_e > \pi/2$. By contrast, the normal components of the capillary forces, which are proportional to $\gamma \sin \theta_e$, have a destabilizing effect for arbitrary $\theta_e > 0$. The underlying physical mechanism can be inferred from the inset of Fig.~\ref{fig:analysis}: a small perturbation of the plate inclination induces a longer moment arm one one side of the plate, such that the resultant torque is destabilizing. Alternatively, one may consider a free energy argument, showing that the immersed state of the wire is energetically unfavorable whenever the liquid is partially wetting (Fig.~\ref{fig:energy}). We found that the dimensionless number associated to this bending mechanism does not only involve the elastocapillary length $\ell_{EC}$, but also the capillary length $\ell_\gamma$. The capillary length appears as it sets the typical moment arm for the torque. Changing the solid from a plate to a thin wire of radius $R \ll \ell_c$, as in Fig.~\ref{fig:bending}, the length scale for the moment arm becomes $R$. By estimating the physical parameters for the wire in Fig.~\ref{fig:bending}, we conclude that in this particular example the instability is not triggered by bending or buckling individually, but by a combination of the two mechanisms.

The analysis of the present paper focussed on the transition of an immersed plate, partially wetted on both sides, to a state where the plate is pushed to free surface. This can be considered as the inverse of the ``piercing'' problem that is relevant e.g. for water striders~\cite{ParkJFM}. In the case of piercing, the contact line remains pinned on the edge of the solid before entering the liquid, such that one side  of the solid remains completely dry. It would be interesting to see if the equivalent of the stability threshold (\ref{eq:onset}) could be derived for piercing as well.

{\bf Acknowledgements:~}We gratefully acknowledge J.~Bico and B.~Roman for enlightening discussions. J.H.S. acknowledges support from Paris-Diderot University where he stayed as an invited professor during the preparation of the manuscript.

%\bibliographystyle{jasanum}

%\bibliography{PlatePRE}
\end{document}